\documentclass[
showpacs,
floatfix,
aps,
prb,
twocolumn,
superscriptaddress,
amssymb,
]{revtex4-1}

\usepackage{times}
\usepackage{amssymb}
\usepackage{latexsym}
\usepackage[dvips]{graphicx}
\usepackage{amsmath}
\usepackage{graphicx}
\usepackage{dcolumn}
\usepackage{amsfonts}
\usepackage{bm}
\usepackage{epsfig}

\newcommand{\be}{\begin{equation}}
\newcommand{\ee}{\end{equation}}
\newcommand{\bea}{\begin{eqnarray}}
\newcommand{\eea}{\end{eqnarray}}

\def\oc{\omega_{\mbox{\scriptsize {c}}}}

\def\tpi{\tau_{\pi}}

\def\ttr{\tau}

\def\rk{R_K}
\def\lp{\left (}
\def\rp{\right )}

\def\rmin{\rho_{\min}}
\def\tl{\tau_L}
\def\ts{\tau_S}

\newcommand{\req}[1]{Eq.\,(\ref{#1})}
\newcommand{\rEq}[1]{Equation\,(\ref{#1})}

\newcommand{\rfig}[1]{Fig.\,\ref{#1}}
\newcommand{\rFig}[1]{Figure \,\ref{#1}}

\newcommand{\rref}[1]{Ref.\,\onlinecite{#1}}
\newcommand{\rrefs}[2]{Refs.\,\onlinecite{#1},\,\onlinecite{#2}}
\newcommand{\rrrefs}[3]{Refs.\,\onlinecite{#1},\,\onlinecite{#2},\,\onlinecite{#3}}

\begin{document}
\title{
Giant negative magnetoresistance in high-mobility 2D electron systems
}
\author{A.\,T. Hatke}
\affiliation{School of Physics and Astronomy, University of Minnesota, Minneapolis, Minnesota 55455, USA}

\author{M.\,A. Zudov}
\email[Corresponding author: ]{zudov@physics.umn.edu}
\affiliation{School of Physics and Astronomy, University of Minnesota, Minneapolis, Minnesota 55455, USA}

\author{J.\,L. Reno}
\affiliation{Sandia National Laboratories, Albuquerque, New Mexico 87185, USA}

\author{L.\,N. Pfeiffer}
\affiliation{Department of Electrical Engineering, Princeton University, Princeton, New Jersey 08544, USA}

\author{K.\,W. West}
\affiliation{Department of Electrical Engineering, Princeton University, Princeton, New Jersey 08544, USA}

\begin{abstract}
We report on a giant negative magnetoresistance in very high mobility GaAs/AlGaAs heterostructures and quantum wells.
The effect is the strongest at $B \simeq 1$ kG, where the magnetoresistivity develops a minimum emerging at $T \lesssim 2$ K. 
Unlike the zero-field resistivity which saturates at $T \simeq 2 $ K, the resistivity at this minimum continues to drop at an accelerated rate to much lower temperatures and becomes {\em several times smaller} than the zero-field resistivity.
Unexpectedly, we also find that the effect is destroyed not only by increasing temperature but also by modest in-plane magnetic fields.
The analysis shows that giant negative magnetoresistance {\em cannot} be explained by existing theories considering interaction-induced or disorder-induced corrections. 
\end{abstract}
\maketitle

Over the past decade, low field magnetotransport in high mobility two-dimensional electron systems (2DESs) became a subject of considerable interest, in part, owing to the discovery of many unexpected phenomena.\citep{zudov:2001a,zudov:2001b,yang:2002,mani:2002,zudov:2003,yang:2003,kukushkin:2004,bykov:2007,zhang:2007c,hatke:2008b,khodas:2010,hatke:2010a,dai:2010,hatke:2011b}
While the characteristic features of the majority of these phenomena are now understood reasonably well, \citep{ryzhii:1970,durst:2003,lei:2003,dmitriev:2003,vavilov:2004,dmitriev:2005,dmitriev:2007,vavilov:2007,khodas:2008,dmitriev:2009b,raichev:2009,dmitriev:2010b,hatke:2011e} there are still exist many unsolved puzzles.
One such puzzle is the recently reported giant microwave photoresistivity peak which emerges in the vicinity of the second harmonic of the cyclotron resonance.\citep{dai:2010,hatke:2011b,hatke:2011c}
While its origin remains unclear, this peak so far has been observed only in 2DESs which also exhibit giant negative magnetoresistance (GNMR).\citep{dai:2010,hatke:2011b}
Therefore, investigating the GNMR effect\citep{note:1} is not only interesting and important in its own right but may also provide necessary clues to account for other phenomena.

The magnetoresistance can be characterized by the ratio $\rho(B)/\rho_0$, where $\rho(B)$ and $\rho_0$ are the longitudinal resistivities measured with and without perpendicular magnetic field $B$, respectively.
In the present study, we focus on the regime of weak magnetic fields where Shubnikov-de Haas oscillations are not yet developed.
In this regime, the characteristic feature of $\rho(B)$ is a broad minimum occurring at $B_0\simeq 1$ kG. 
Quite remarkably, the resistivity at this minimum, $\rho(B_0)\equiv\rmin$, can be significantly lower than $\rho_0$, i.e. $\rmin/\rho_0\ll 1$, in very high mobility samples.\citep{dai:2010,hatke:2011b}
In what follows we will use the value of $\rmin/\rho_0$ to quantitatively describe the GNMR.

While negative magnetoresistance effect has been known for nearly three decades,\citep{paalanen:1983,choi:1986,li:2003} systematic experimental studies in very high mobility ($\mu \sim 10^7$ cm$^2$/Vs) 2DESs have appeared only recently.
More specifically, \textcite{bockhorn:2011} reported that the effect quickly disappears with increasing density; $\rmin/\rho_0$ increased from $\approx 0.3$ to $\approx 0.7$ as the carrier density changed from $\approx 2$ to $\approx 3\cdot 10^{11}$ cm$^{-2}$.\citep{note:4}
In addition, it was found\citep{bockhorn:2011} (for the carrier density of $\approx 2.3 \cdot 10^{11}$ cm$^{-2}$) that the minimum resistivity roughly doubles when the temperature is raised from 0.1 to 0.8 K.

In this Rapid Communication we systematically investigate the roles of temperature and in-plane magnetic field on the GNMR effect observed in high mobility GaAs/AlGaAs heterostructures and quantum wells.
In all of our samples, the effect manifests itself as a well defined minimum in the longitudinal resistivity emerging at $B_0 \simeq 1$ kG.
At low temperatures and low in-plane fields, the resistivity at this minimum is {\em a small fraction} of the zero-field resistivity. 
Remarkably, the GNMR is quickly suppressed not only by temperature but also by modest (a few kG) in-plane magnetic fields.
Our analysis of the low-field magnetoresistivity shows that the observed GNMR {\em cannot} be explained by existing theories considering either interaction-induced or disorder-induced corrections to the Drude resistivity.

Our samples (A, B, and C) are lithographically defined Hall bars (widths $w_{\rm A}=50$ $\mu$m, $w_{\rm B}=150$ $\mu$m, $w_{\rm C}=100$ $\mu$m).
Sample A is fabricated from a GaAs/AlGaAs Sandia-grown heterostructure with density $n_{\rm A} \approx 1.6\cdot10^{11}$ cm$^{-2}$ and mobility $\mu_{\rm A} \approx 5.4 \cdot 10^6$ cm$^2$/Vs.
Sample B (C) is made from a Princeton-grown 24(30) nm-wide GaAs/AlGaAs quantum well with density $n_{\rm B} \approx 4.3\cdot 10^{11}$ cm$^{-2}$ ($n_{\rm C} \approx 3.4\cdot10^{11}$) and mobility $\mu_{\rm B} \approx 1.0 \cdot 10^7$ cm$^2$/Vs ($\mu_{\rm C} \approx 1.2 \cdot 10^7$ cm$^2$/Vs). 
Magnetoresistivity $\rho(B)$ was measured in a $^{3}$He cryostat at temperatures up to $T = 6.0$ K using a standard low frequency lock-in technique.

\begin{figure}[t]
\includegraphics{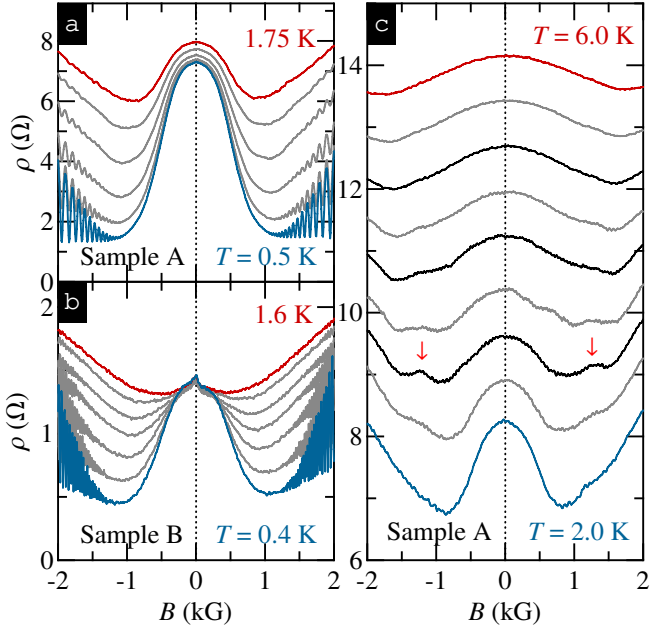}
\vspace{-0.1 in}
\caption{(Color online)
(a) $\rho(B)$ of sample A at $0.5$ K $ \le T \le 1.75$ K, with a step $\Delta T = 0.25$ K.
(b) $\rho(B)$ of sample B at $0.4$ K $ \le T \le 1.6$ K, $\Delta T = 0.2$ K.
(c) $\rho(B)$ of sample A at $2$ K $ \le T \le 6$ K, $\Delta T = 0.5$ K. 
Arrows mark the second order maxima of phonon-induced resistance oscillations (see text).
}
\vspace{-0.15 in}
\label{fig1}
\end{figure}
In \rfig{fig1}(a) [(b)] we present the magnetoresistivity $\rho(B)$ in sample A [sample B] measured at $T$ from 0.5 K to 1.75 K [from 0.4 K to 1.6 K], in a step of 0.25 K [0.2 K].
In addition to Shubnikov-de Haas oscillations, both samples reveal a GNMR effect marked by a pronounced minimum which occurs at $B_0 \simeq 1$ kG and becomes progressively deeper with decreasing $T$;
in contrast to the zero-field resistivity, $\rho_0$, which remains nearly temperature-independent, the resistance at this minimum, $\rmin$, decays rapidly and becomes a small fraction of the zero-field resistivity. 
For example, in sample A, $\rmin/\rho_0 \approx 0.2$ at $T = 0.5$ K.

To examine the MR effect at higher $T$, we present in \rfig{fig1}(c) the magnetoresistivity $\rho(B)$ in sample A  at temperatures from 2 K to 6 K, in a step of 0.5 K. 
Here, we notice that at $T < 4\,{\rm K}$, $\rho(B)$ exhibits phonon-induced resistance oscillations, owing to resonant electron scattering on thermally excited $2k_F$-acoustic phonons.\citep{zudov:2001b,zhang:2008,hatke:2009b,raichev:2009,dmitriev:2010b,hatke:2011d}
The second order maxima of these oscillations occur at $B \approx 1.3 $ kG, as marked by $\downarrow$ next to the trace at $T = 3.0$ K  in \rfig{fig1}(c).\citep{note:5}
At $T \gtrsim 4$ K, the position of the resistivity minimum is shifted to a higher field ($\approx 1.5$ kG) and both $\rho_0$ and $\rmin$ grow at about the same rate, as evidenced by roughly parallel traces in \rfig{fig1}(c). 
The spacing between adjacent traces remains roughly constant indicating linear temperature dependence of the resistivity over the entire range of magnetic fields.

\begin{figure}[t]
\includegraphics{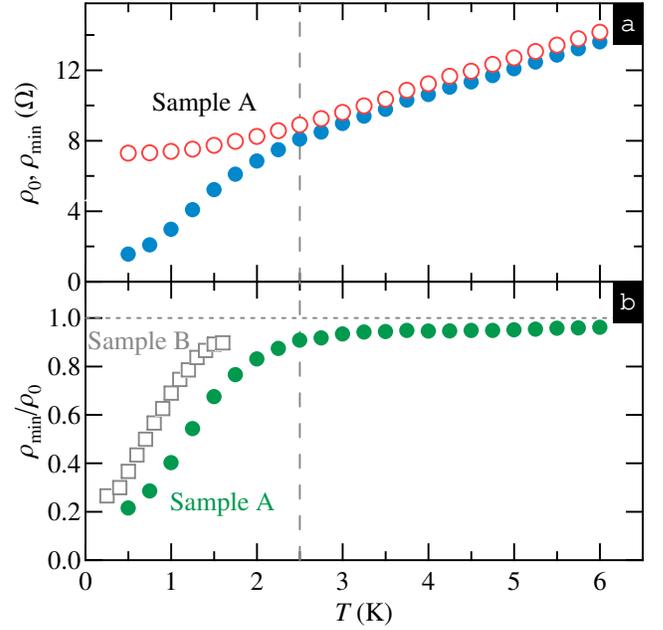}
\vspace{-0.1 in}
\caption{(Color online)
(a) $\rho_0$ (open circles) and $\rmin$ (solid circles) versus $T$ in sample A.
Vertical line ``separates'' high and low temperature regimes in sample A.
(b) $\rmin/\rho_0$ versus $T$ in sample A (circles) and in sample B (squares).
}
\vspace{-0.15 in}
\label{fig2}
\end{figure}
For a quantitative analysis of the GNMR we present in \rfig{fig2}(a) the zero-field resistivity, $\rho_0$ (open circles), and the resistivity at the minimum, $\rmin$ (solid circles), measured in sample A for each $T$ studied.
The data clearly show that at $T \gtrsim 2.5$ K (to the right of the dashed vertical line), the resistivities are close to each other, $\rho_0 \simeq \rmin$, both featuring very similar, approximately linear, temperature dependence. 
Such behavior is consistent with the electron scattering on thermal acoustic phonons.\citep{stormer:1990,hatke:2009b}

At lower temperatures, $T \lesssim 2.5$ K (to the left of the vertical line), the $T$ dependences of $\rho_0$ and $\rmin$ become markedly different.
The decrease of $\rho_0$ gets considerably {\em slower} as the acoustic phonon contribution becomes irrelevant and the resistivity saturates at a value determined by impurity scattering.\citep{stormer:1990,hatke:2009b,note:14}
Quite remarkably, in contrast to $\rho_0$, $\rmin$ not only continues to drop at lower temperatures but also does so at a much {\em faster} rate. 
Such a sudden change of the temperature dependence of $\rmin$ is totally unexpected.
Quantitatively, once the temperature is lowered from 2.5 K to 0.5 K, $\rho_0$ decreases only by about 20\% while $\rmin$ drops by more than a factor of five.\citep{note:13}
 
Using $\rho_0$ and $\rmin$ shown in \rfig{fig2}(a), we calculate $\rmin/\rho_0$ and present the result (circles) in \rfig{fig2}(b) as a function of temperature.
Results for sample B obtained in the same way using the data in \rfig{fig1}(b) are represented by squares.
Both samples show a rapid increase of $\rmin/\rho_0$ with increasing temperature and eventual saturation at $\rmin/\rho_0 \simeq 1$.

\begin{figure}[t]
\includegraphics{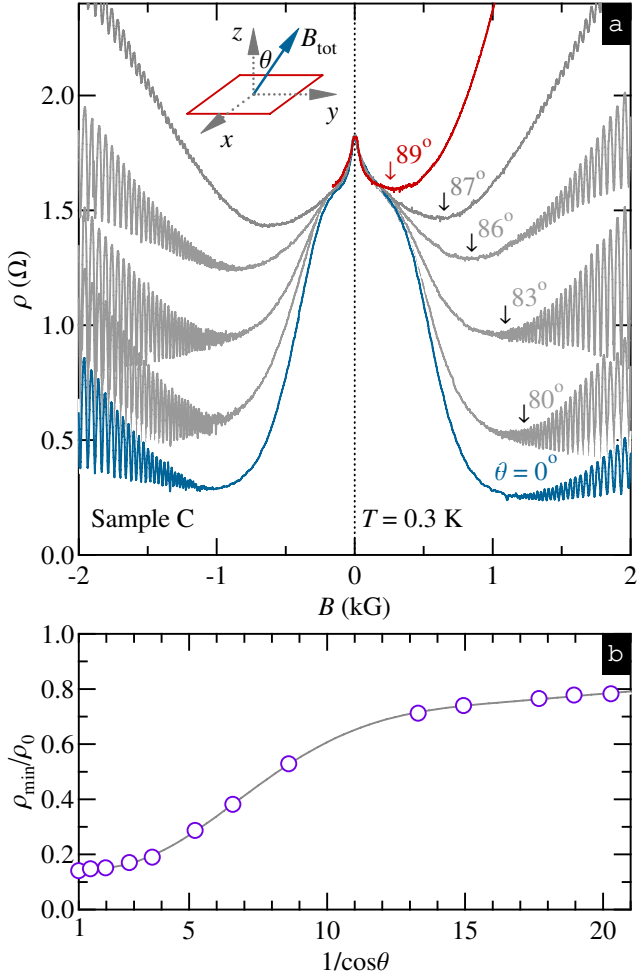}
\vspace{-0.1 in}
\caption{(Color online)
(a) $\rho(B)$ of sample C at $T = 0.3$ K at different tilt angles $\theta$ (as marked).
(b) $\rmin/\rho_0$ versus $1/\cos\theta$ (circles). 
Solid curve is a guide to an eye. 
}
\vspace{-0.15 in}
\label{fig3}
\end{figure}
We next examine the effect of an in-plane magnetic field which is introduced by tilting the sample normal by angle $\theta$ with respect to the magnet axis.
\rFig{fig3}(a) shows magnetoresistivity $\rho(B)$ at selected $\theta$ from $0^\circ$ to $89^\circ$ measured in sample C at $T\simeq 0.3$ K.
At $\theta = 0^\circ$ we again observe GNMR characterized by $\rmin/\rho_0 \approx 0.14$.
With increasing $\theta$ the data reveal rather complex behavior; 
$\rmin$ increases while $B_0$ becomes smaller, decreasing roughly by a factor of four at the highest angle.

To estimate the characteristic in-plane field required to suppress GNMR we extract $\rmin/\rho_0$ from the data in \rfig{fig3}(a) and present the result in \rfig{fig3}(b) as a function of $1/\cos\theta$.
We find that $\rmin/\rho_0$ doubles at $1/\cos\theta\,\simeq\,5$ which gives the scale of the in-plane field, $B_\parallel = B_0/\cos\theta\,\simeq\,5$ kG.
We note that similar in-plane field values were found necessary to suppress microwave-induced\citep{yang:2006} and Hall field-induced\citep{hatke:2011a} resistance oscillations occurring in a similar perpendicular field range. 
At higher tilt angles $\rmin/\rho_0$ appears to saturate at $\approx 0.8$.

At first glance observed increase of $\rmin$ with increasing tilt angle might originate from the in-plane field-induced positive magnetoresistance effect, recently reported in very high mobility 2DEG.\citep{zhou:2010}
However, according to \rref{zhou:2010} {\em an order of magntitude higher} $B_\parallel$ is needed to double the resistance in a 30 nm-wide quantum well.
Therefore, further studies are needed to clarify the origin of the $B_\parallel$-induced suppression of the GNMR effect.

\begin{figure}[t]
\includegraphics{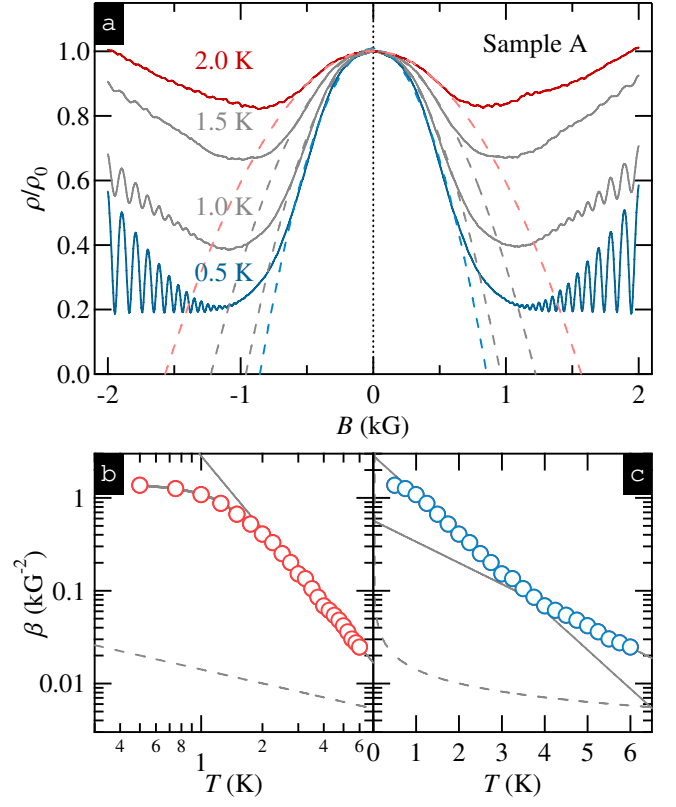}
\vspace{-0.1 in}
\caption{(Color online)
(a) Solid curves represent 
$\rho(B)/\rho_0$ measured in sample A at $T$ from $0.5$ K to $2.0$ K, as marked.
Dashed curves are fits to the data, $\rho(B)/\rho_0=1-\beta B^2$, at $|B| \le 0.5$ kG.
(b,c) $\beta$ versus $T$.
Solid lines are fits to the data (see text) and dashed lines are $\beta_i^{\rm sm}$ calculated using \req{int}. 
}
\vspace{-0.15 in}
\label{fig4}
\end{figure}

In the remainder of this Rapid Communication, we focus on the temperature dependence of the low-field magnetoresistivity preceeding the formation of the deep minimum at $B=B_0$.
More specifically, we analyze the low $B$ part of the data in terms of 
\be
\frac{\rho(B)}{\rho_0} = 1 - \beta B^2\,,
\label{eq.fit}
\ee
and then examine $\beta$ as a function of temperature.
In \rfig{fig4}(a) we plot normalized magnetoresistivity, $\rho(B)/\rho_0$, measured in sample A at $T$ from $0.5$ K to $2.0$ K, in a step of 0.5 K.\citep{note:6}
To extract $\beta$ we fit the data using \req{eq.fit} over the range $|B| \le 0.5$ kG (cf.\,dashed lines) and observe that the curvature of the low field resistivity $\beta$ {\em decreases} with increasing temperature.

After repeating the fitting procedure for all other $T$ studied, we present extracted $\beta$ in \rfig{fig4}(b) and \rfig{fig4}(c) using log-log and log-linear scale, respectively.
First, we notice that at $T \lesssim 1$ K, $\beta$ shows a sign of saturation and can be well described by $\beta \approx 1.45 - T^2/T_0^2$, $T_0 \approx 1.7$ K [cf.\, solid curve in \rfig{fig4}(b)].
At higher $T$ the data can be described by either $\beta \propto T^{-2.6}$, $T\gtrsim 2.5$ K [cf. solid line in \rfig{fig4}(b)] or by $\beta \propto \exp(-T/T_{1,2})$, where $T_1 \approx 1.0~{\rm K}$ for $1.0~{\rm K}\lesssim T\lesssim 3.5~{\rm K}$ and $T_2 \approx 1.9~{\rm K}$ for $3.5~{\rm K}\lesssim T \lesssim 6.0~{\rm K}$ [cf. solid lines in \rfig{fig4}(c)].
It is clear that the temperature dependence of $\beta$ is rather complex which is likely a result of one or several crossovers between different regimes. 
In what follows we examine $\beta(T)$ in terms of existing theoretical models and compare the results of our analysis to other experimental studies.\citep{dai:2010,bockhorn:2011}

{\em Quasiclassical disorder model},\citep{mirlin:2001} predicts a parabolic negative magnetoresistance, see \req{eq.fit}, with $\beta$ given by
\be
\beta_d = \frac {e^2} {2\pi n_S p_F^2} \lp \frac \tl {2\ts} \rp^{1/2}\,,~~~0 < \tl^{-1} \ll \ts^{-1}\,.
\label{dis}
\ee
Here, $\tl^{-1}$ and $\ts^{-1}$ are long- and short-range disorder momentum relaxation rates, $\ttr^{-1}=\tl^{-1}+\ts^{-1}$,\citep{note:2} 
$n_S$ is the areal density of short-range scatterers, and $p_F$ is the Fermi momentum.
\rEq{dis} is valid for $\beta_d B ^2 \ll 1$ and at higher $B$ the resistivity is expected to saturate at 
$\rmin \simeq \rho_0\cdot(\ts/\tl) \ll \rho_0$.\citep{mirlin:2001,polyakov:2001}

While the {\em disorder model} can, in principle, lead to GNMR, it clearly fails to explain our experimental findings.
First, as shown above, $\beta$ exhibits strong dependence on temperature which does not enter \req{dis}.
Second, we believe that the assumption of $\tl^{-1} \ll \ts^{-1}$ is not satisfied in our samples.
Indeed, the analysis of Hall field-induced resistance oscillations in sample A\citep{note:7} suggests opposite relation, $\tl^{-1} \simeq 5 \ts^{-1}$.
We finally notice that while \rref{dai:2010} concluded that the MR in their samples can be consistently described by \req{dis},\citep{note:3} neither the temperature dependence nor the validity of $\tl^{-1} \ll \ts^{-1}$ condition has been examined.

{\em Electon-electron interaction model},\citep{gornyi:2003,li:2003,gornyi:2004} on the other hand, predicts a temperature-dependent magnetoresistance.
In the ballistic regime, $\hbar/\ttr \ll k_BT$, and for smooth disorder potential this model also leads to \req{eq.fit}, with $\beta$ given by\citep{gornyi:2004}
\be
\beta_i^{\rm sm} = \mu^2 \frac {\rho_0}{\rk} \frac {c_0} \pi \lp \frac {\hbar/\ttr} {k_B T} \rp^{1/2},~~~\ts^{-1} = 0\,.
\label{int}
\ee
Here, $\rk=h/e^2$ is the von Klitzing constant and $c_0 = 3\zeta(3/2)/16\sqrt{\pi} \simeq 0.276$. 
However, \req{int} also fails to describe our findings.
Indeed, taking $T=1$ K as an example, our experiment gives $\beta \approx 1.1$ kG$^{-2}$ which is nearly {\em two orders of magnitude larger} than $\beta_i^{\rm sm}\approx 0.014$ kG$^{-2}$ obtained from \req{int}.
Comparison of $\beta_i^{sm}$ obtained using \req{int} [cf.\, dashed line in Figs. \ref{fig4}(b) and \ref{fig4}(c)] with our data shows that the discrepancy remains significant over the whole range of $T$ studied.
Moreover, it this clear that the interaction model fails to explain our data even on a qualitative level.
We also notice that significant disagreement with \req{int} was found in \rref{bockhorn:2011} reporting low-temperature $\beta$ which is roughly $30$ ($n \approx 2 \cdot 10^{11}$ cm$^{-2}$) to $150$ ($n \approx 3 \cdot 10^{11}$ cm$^{-2}$) times larger than $\beta_i^{\rm sm}$.\citep{note:12}

We next consider several scenarios for the observed discrepancy.
First, in a realistic high-mobility 2DEG, sharp disorder, which is not present in \req{int}, plays a crucial role in many of the low-field magnetotransport phenomena.\citep{mirlin:2001,polyakov:2001,yang:2002,zhang:2007a,zhang:2007c,zhang:2008,hatke:2008a,vavilov:2007,hatke:2008b,hatke:2009a,hatke:2009c,dmitriev:2009b,khodas:2008,khodas:2010,hatke:2010b} 
For the case of mixed disorder potential \req{int} is generalized to\citep{gornyi:2004}
\be
\beta_i^{\rm mix} = \lp 4 - \frac {3\ttr}{\tl} \rp \sqrt{\frac \tl \ttr} \beta_i^{\rm sm}\,.
\label{int2}
\ee
If $\tl^{-1} \ll \ts^{-1}$, there appears a parametrically large factor $4(\tl/\ttr)^{1/2} \gg 1$ which leads to $\beta_i^{\rm mix} \gg \beta_i^{\rm sm}$.
However, in our sample A, as mentioned above, $\tl^{-1} \simeq 5\ts^{-1}$ from which we estimate $(4-3\ttr/\tl)\sqrt{\tl/\ttr}\approx 1.5$.
Such a small factor is clearly not sufficient to explain the discrepancy.

Another possible cause for large $\beta$ is the disorder-induced $T$-independent correction, similar to that given by \req{dis}. 
Assuming that the contributions are additive, one has $\beta = \beta_d + \beta_i$, where $\beta_d\,(\beta_i)\propto T^0\,(T^{-1/2})$. 
It is clear, however, that the experimentally obtained $\beta(T)$ cannot be described by such dependence.\citep{note:8}

Finally, theory should consider a possibility that the low-temperature magnetoresistance originates primarily from the quasiclassical disorder mechanism which, however, is significantly altered by the electron-electron interactions with increasing temperature.\citep{note:11}
However, such a theory remains a subject of future work.

In summary, a giant negative magnetoresistance effect in high-mobility GaAs/AlGaAs heterostructures and quantum wells is marked by a pronounced minimum of the longitudinal resistivity appearing at $B \simeq 1$ kG.
The temperature dependence clearly reveals a crossover between two distinct regimes.
In the high temperature regime, the zero-field resistivity and the minimum resistivity both exhibit linear temperature dependence, due to scattering on thermal acoustic phonons.
In the low temperature regime, however, zero-field resistivity quickly saturates but the minimum resistivity continues to decrease at an even faster rate eventually becoming a small fraction of the zero-field resitivity.
Unexpectedly, we also find that the GNMR is destroyed not only by temperature but also by very modest (a few kG) in-plane magnetic fields. 
Finally, our analysis of the low-field magnetoresistivity demonstrates that the GNMR effect cannot be understood by existing theoretical models considering either interaction-induced or disorder-induced corrections, even on a qualitative level. 
Taken together, these findings provide important clues for emerging theories and should help to elucidate the origin of the GNMR in very high mobility 2DES. 

We thank M. Dyakonov, R. Houg, M. Khodas, D. Polyakov, M. Raikh, and B. Shklovskii for discussions and 
G. Jones, T. Murphy, and D. Smirnov for technical assistance.
A portion of this work was performed at the 
NHMFL, which is supported by NSF Cooperative Agreement No. DMR-0654118, by the State of Florida, and by the DOE and at the Center for Integrated Nanotechnologies, a U.S. Department of Energy, Office of Basic Energy Sciences user facility and at the Center for Integrated Nanotechnologies, a U.S. Department of Energy, Office of Basic Energy Sciences user facility.   
The work at Minnesota was supported by the NSF Grant No. DMR-0548014 (measurements at Minnesota on Samples A and B) and by the DOE Grant No. DE-SC002567 (tilt-field measurements at NHMFL on sample C). 
The work at Princeton was partially funded by the Gordon and Betty Moore Foundation and the NSF MRSEC Program through the Princeton Center for Complex Materials (DMR-0819860) and the work at Sandia was supported by the Sandia Corporation under Contract No. DE-AC04-94AL85000.
Sandia National Laboratories is a multi-program laboratory managed and operated by Sandia Corporation, a wholly owned subsidiary of Lockheed Martin Corporation, for the U.S. Department of Energy's National Nuclear Security Administration under contract DE-AC04-94AL85000.


\begin{thebibliography}{68}
\bibitem[{not({\natexlab{b}})}]{note:4}
\bibinfo{note}{The corresponding mobility variation was from $\mu \simeq
  0.9\cdot10^7$ cm$^2$/Vs to $\mu \simeq 1.1\cdot10^7$ cm$^2$/Vs.}
\bibitem[{not({\natexlab{f}})}]{note:6}
\bibinfo{note}{In order to improve the fitting procedure, a slight asymmetry of
  the raw data in \rfig{fig1}\,(a) was eliminated by subtracting a linear in
  $B$ component to ensure $\rho(B_0)=\rho(-B_0)$.}
\bibitem[{not({\natexlab{j}})}]{note:12}
\bibinfo{note}{The disagreement with \req{int} is not only in the magnitude of
  $\beta$ but also in its much stronger dependence on density.}
\end{thebibliography}

\end{document}